\pdfoutput=1
\documentclass[sigconf, screen=true]{acmart}
\usepackage[T1]{fontenc}
\usepackage[utf8]{inputenc}

\definecolor{bostonuniversityred}{rgb}{0.8, 0.0, 0.0}
\newcommand{\hk}[1]{{#1}} % Hedvig
\newcommand{\rn}[1]{#1} % Rajmund
\usepackage{subfig}
\usepackage{float}
\usepackage[capitalise, noabbrev]{cleveref} % Automatically include e.g. "Section" in \ref
\usepackage{soul} % strikethrough text
\usepackage{colortbl} % coloured tables

\definecolor{lightgray}{gray}{0.925}
\usepackage{multirow}
\usepackage{array}
\usepackage{tabularx}
\usepackage{longtable}
\usepackage{pdflscape}
\usepackage{pifont}
\usepackage{dblfloatfix}
\usepackage{arydshln}
\usepackage{placeins}
\usepackage{enumitem} % Better control of lists and their spacing
\setlist{nolistsep} % R

\newcommand\blfootnote[1]{%
  \begingroup
  \renewcommand\thefootnote{}\footnote{#1}%
  \addtocounter{footnote}{-1}%
  \endgroup
}

\AtBeginDocument{%
  \providecommand\BibTeX{{%
    \normalfont B\kern-0.5em{\scshape i\kern-0.25em b}\kern-0.8em\TeX}}}

\setcopyright{rightsretained}

\begin{document}

\title{Speech2Properties2Gestures: Gesture-Property Prediction as a Tool for Generating Representational Gestures from Speech}
%\title{Speech2Properties2Gestures: Generating representational gestures based on predicting gesture properties from speech}


\author{Taras Kucherenko}
\authornote{Robotics, Perception, and Learning, KTH Royal Institute of Technology, Sweden.}
\affiliation{%
  \institution{}
  \city{}
  \state{}
  \country{}
}
%\affiliation{KTH Royal Institute of Technology}
\email{tarask@kth.se}

\author{Rajmund Nagy}
\authornote{Speech, Music, and Hearing, KTH Royal Institute of Technology, Sweden.}
\affiliation{%
  \institution{}
  \city{}
  \state{}
  \country{}
}
%\affiliation{KTH Royal Institute of Technology}
\email{rajmundn@kth.se}

\author{Patrik Jonell}
\authornotemark[2]
\affiliation{%
  \institution{}
  \city{}
  \state{}
  \country{}
}
%\affiliation{KTH Royal Institute of Technology}
\email{pjjonell@kth.se}

\author{Michael Neff}
\authornote{University of California, Davis, United States.}
%\authornotemark[2]
\affiliation{%
  \institution{}
  \city{}
  \state{}
  \country{}
}
\email{mpneff@ucdavis.edu}

\author{Hedvig Kjellström}
\authornotemark[1]
\affiliation{%
  \institution{}
  \city{}
  \state{}
  \country{}
}
%\affiliation{KTH Royal Institute of Technology}
\email{hedvig@kth.se}

\author{Gustav Eje Henter}
\authornotemark[2]
\affiliation{%
  \institution{}
  \city{}
  \state{}
  \country{}
}
%\affiliation{KTH Royal Institute of Technology}
\email{ghe@kth.se}

%\affiliation{%
%  \institution{The Th{\o}rv{\"a}ld Group}
%\email{larst@affiliation.org}

%\renewcommand{\shortauthors}{Kucherenko, et al.}
%\renewcommand{\shortauthors}{Kucherenko, Nagy, Jonell, Neff, Kjellström, and Henter}

\copyrightyear{2021}
\acmYear{2021}
\acmConference[IVA '21]{21th ACM International Conference on Intelligent Virtual Agents}{September 14--17, 2021}{Virtual Event, Japan}
\acmBooktitle{21th ACM International Conference on Intelligent Virtual Agents (IVA '21), September 14--17, 2021, Virtual Event, Japan}\acmDOI{10.1145/3472306.3478333}
\acmISBN{978-1-4503-8619-7/21/09}

\begin{abstract}
We propose a new framework for gesture generation, aiming to allow data-driven approaches to produce more semantically rich gestures. Our approach first predicts whether to gesture, followed by a prediction of the gesture properties. Those properties are then used as conditioning for a modern probabilistic gesture-generation model capable of high-quality output. This empowers the approach to generate gestures that are both diverse and representational. Follow-ups and more information can be found on the project page:\\ \href{https://svito-zar.github.io/speech2properties2gestures/}{https://svito-zar.github.io/speech2properties2gestures/}
%In this way, the model is capable of predicting gestures that are both diverse and representational.
\end{abstract}




\begin{teaserfigure}
\centering
  \vspace{-3.5ex}
  \includegraphics[width=.69\textwidth]{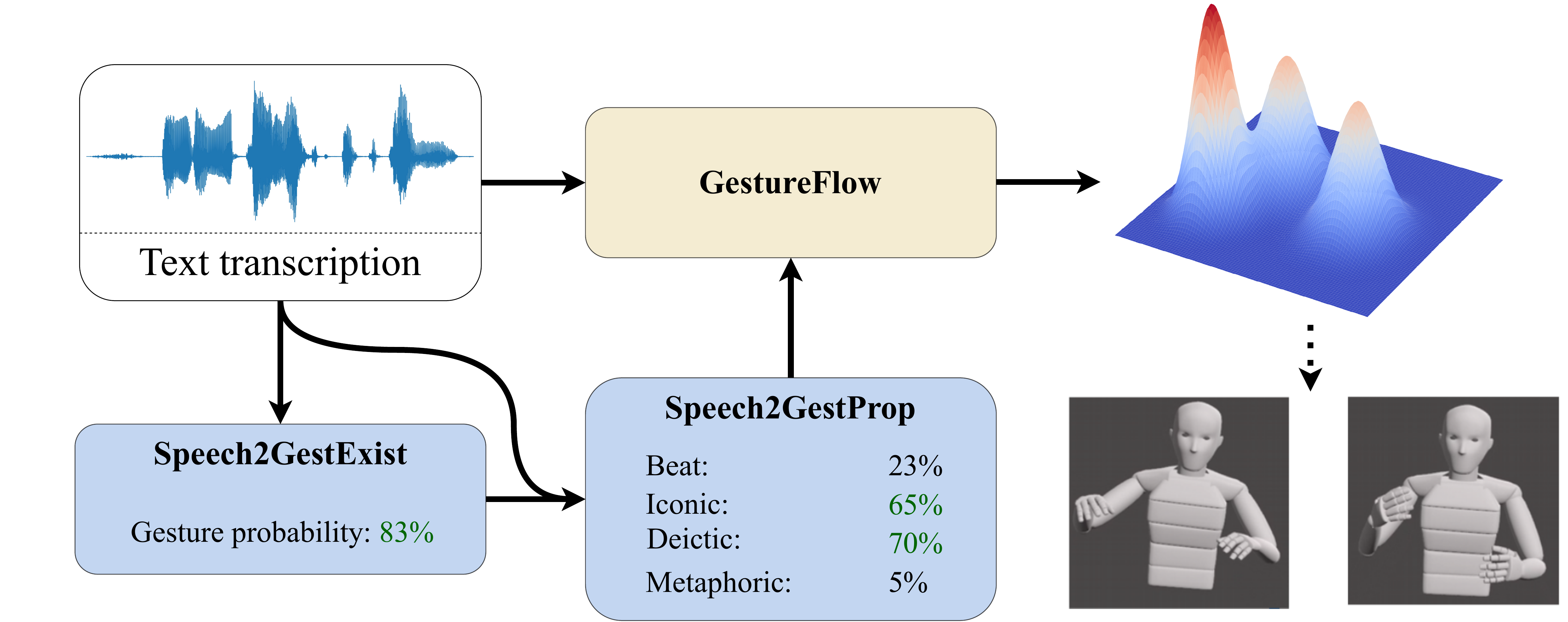}
  \caption{Overview of the proposed framework. We first use speech text and audio to predict whether or not the agent should gesture. After that, we predict several gesture properties, such as gesture type. Finally, gestures are generated by a probabilistic model (e.g., a normalizing flow) conditioned on text, audio, and predicted gesture properties together.}
  \Description{A box diagram with arrows showing how speech audio and text transcription are fed into a Speech2GestExist to decide when to generate a gesture and, if yes, also are fed into Speech2GestProp to predict the properties of the gesture to be generated. The predicted properties, together with audio and text, all feed into a box marked GestureFlow, which outputs a distribution over pose sequences for an avatar depicted in the figure.}
  \label{fig:teaser}
\end{teaserfigure}

\settopmatter{printfolios=true} % Turn this on for page numbering on arXiv
\maketitle

\section{Introduction and Background}
%\section{Why you should care}
A large part of human communication is non-verbal \cite{knapp2013nonverbal} and often takes place through co-speech gestures %is an important part of non-verbal communication, since it communicate a large share of non-verbal content 
\cite{mcneill1992hand, kendon2004gesture}. Co-speech gesture behavior in embodied agents has been shown to %For example, agent's gestures can 
help with learning tasks \cite{bergmann2013virtual} and %can 
lead to  greater emotional response \cite{wu2014effects}. Gesture generation is hence an important part of both automated character animation and human-agent interaction. 

%\section{Previous work}
Early dominance of rule-based approaches \cite{cassell1994animated, kopp2004synthesizing, ng2010synchronized,marsella2013virtual} has been challenged by data-driven gesture generation systems \cite{neff2008gesture, bergmann2009GNetIc, yoon2018robots, kucherenko2020gesticulator, yoon2020speech, ahuja2020no, ferstl2020understanding}. These latter systems first only considered a single speech modality (either audio or text) \cite{neff2008gesture, bergmann2009GNetIc, yoon2018robots, kucherenko2021moving}, but are now shifting to use both %of the 
audio and text together \cite{kucherenko2020gesticulator, yoon2020speech, ahuja2020no}.

While rule-based systems provide control over the communicative function of output gestures, they lack variability and require much manual effort to design. Data-driven systems, on the other hand, need less manual work and are very flexible, but most existing systems do not provide much control over communicative function and generated gestures have little relation to speech content \cite{kucherenko2020genea}. 

\definecolor{lightgray}{gray}{0.925}

\begin{table*}[t!]
\centering
%\rowcolors{3}{}{lightgray} % Coloured rows
% Better-than-chance formatting
\newcommand{\highlight}[1]{\textbf{\textcolor{Maroon}{#1}}}
% Make the table fit the whole page
\resizebox{\textwidth}{!}{
% Taller lines
\renewcommand{\arraystretch}{1.5}

%One right-aligned column followed by 13*3=39 columns (e.g. deictic is 3 columns because I align on the plus-minus by putting it into a separate middle column). The column triplets are right-center-left aligned.
\begin{tabular}{@{\kern\tabcolsep}r *{39}{r@{\hspace{\tabcolsep}}c@{\hspace{\tabcolsep}}l}@{\kern\tabcolsep}}

\toprule
& \multicolumn{12}{c}{\textbf{Gesture category} [Macro $\text{F}_1$]} & \multicolumn{12}{c}{\textbf{Gesture semantics} [Macro $\text{F}_1$]} & \multicolumn{15}{c}{\textbf{Gesture phase} [$\text{F}_1$]} \\

\cmidrule(lr){2-13}
\cmidrule(lr){14-25}
\cmidrule(lr){26-40}

Label & \multicolumn{3}{c}{deictic} & \multicolumn{3}{c}{beat} & \multicolumn{3}{c}{iconic} & \multicolumn{3}{c}{discourse} & \multicolumn{3}{c}{amount} & \multicolumn{3}{c}{shape} & \multicolumn{3}{c}{direction} & \multicolumn{3}{c}{size} & \multicolumn{3}{c}{pre-hold} & \multicolumn{3}{c}{post-hold} & \multicolumn{3}{c}{stroke} & \multicolumn{3}{c}{retr.} & \multicolumn{3}{c}{prep.} \\

Relative frequency & \multicolumn{3}{c}{29.05\%} & \multicolumn{3}{c}{14.47\%} & \multicolumn{3}{c}{72.03\%} & \multicolumn{3}{c}{12.78\%} & \multicolumn{3}{c}{4.7\%} & \multicolumn{3}{c}{13.1\%} & \multicolumn{3}{c}{13.7\%} & \multicolumn{3}{c}{1.9\%} & \multicolumn{3}{c}{0.6\%} & \multicolumn{3}{c}{12.2\%} & \multicolumn{3}{c}{40.9\%} & \multicolumn{3}{c}{14.8\%} & \multicolumn{3}{c}{30.8\%} \\ 

\specialrule{\lightrulewidth}{0pt}{0pt}

RandomGuess  & 50\% & $\pm$ & 2\%  & 50\% & $\pm$ & 2\% & 50\% & $\pm$ & 1.5\%        & 50\% & $\pm$ & \hphantom{0}2\%  & 49\% & $\pm$ & 1\%   & 49\% & $\pm$ & 2\%  & 49\% & $\pm$ & 2\%  & 50\% & $\pm$ & 1\%  & 1.3\% & $\pm$ & 4\%   & 12\% & $\pm$ & 4\%  & 42\% & $\pm$ & 4\%   & 14\% & $\pm$ & 5\% & 30\% & $\pm$ & 3\%          \\ 

\specialrule{\lightrulewidth}{0pt}{0pt}
\rowcolor{white}
%AudioOnly       & 50\% & $\pm$ & 3\%  & 46\% & $\pm$ & 2\% & 53\% & $\pm$ & 5\%  & 50\% & $\pm$ & \hphantom{0}2\%  & 49\% & $\pm$ & 2\%   & 49\% & $\pm$ & 4\%  & 50\% & $\pm$ & 2\%  & 49\% & $\pm$ & 1\%  & & 0\% &    & 5\% & $\pm$ & 3\%   & \highlight{53\%} & \highlight{$\pm$} & \highlight{8\%}  & 12\% & $\pm$ & 5\% & \highlight{40\%} & \highlight{$\pm$} & \highlight{4\%} \\

\rowcolor{lightgray}

%TextOnly        & \highlight{60\%} & \highlight{$\pm$} & \highlight{6\%} & 54\% & $\pm$ & 3\% & \highlight{64\%} & \highlight{$\pm$} & \highlight{7\%} & \highlight{58\%} & \highlight{$\pm$} & \highlight{\hphantom{0}8\%} & \highlight{62\%} & \highlight{$\pm$} & \highlight{11\%} & \highlight{65\%} & \highlight{$\pm$} & \highlight{8\%} & \highlight{60\%} & \highlight{$\pm$} & \highlight{7\%} & 57\% & $\pm$ & 9\%  & 1\% & $\pm$ & 2\%     & 21\% & $\pm$ & 12\% & \highlight{51\%} & \highlight{$\pm$} & \highlight{12\%} & 25\% & $\pm$ & 9\% & \highlight{42\%} & \highlight{$\pm$} & \highlight{7\%} \\

\rowcolor{white}

ProposedModel  & \highlight{60\%} & \highlight{$\pm$} & \highlight{6\%} & 53\% & $\pm$ & 6\% & \highlight{63\%} & \highlight{$\pm$} & \highlight{5\%} & \highlight{59\%} & \highlight{$\pm$} & \highlight{\hphantom{0}7\%} & \highlight{63\%} & \highlight{$\pm$} & \highlight{8\%}  & \highlight{65\%} & \highlight{$\pm$} & \highlight{6\%} & \highlight{62\%} & \highlight{$\pm$} & \highlight{8\%} & \highlight{59\%} & \highlight{$\pm$} & \highlight{9\%} & 0.5\% & $\pm$ & 1.3\% & 23\% & $\pm$ & 12\% & \highlight{47\%} & \highlight{${\pm}$} & \highlight{10\%} & 25\% & $\pm$ & 5\% & \highlight{45\%} & \highlight{$\pm$} & \highlight{6\%} \\ 

\bottomrule
\end{tabular}}

\caption{Gesture-property prediction scores for random guessing and our trained predictors using text and audio modalities together. \highlight{Bold, coloured numbers} indicate that the label in question can be predicted better than chance.\vspace{-3ex}}
\label{tab:exp_results}

\end{table*}
%\renewcommand{\arraystretch}{1}
%\normalsize

This paper continues recent efforts to bridge the gap between the two paradigms \cite{ferstl2020understanding, saund2021cmcf, yunus2020sequence}. The most similar prior work is \citeauthor{yunus2020sequence}~\cite{yunus2020sequence} where %they predict where to place gestures 
gesture timing and duration were predicted
based %only 
on acoustic features only. %We are 
The method proposed here differs from %them 
their approach in three ways: 1) it considers % we consider 
not only audio but also text as input; 2) it models %we model 
not only gesture phase, but multiple gesture properties; 3) it also provides %we also propose 
a framework for integrating these gesture properties in a data-driven gesture-generation system.
%Our data-driven system, depicted in Figure \ref{fig:teaser}, both predict and use various gesture properties. 

The \hk{proposed} approach helps decouple different aspects of gesticulation and can leverage database information about gesture timing and content with modern, high-quality data-driven animation.

\section{Proposed Model}
Our unified model uses speech text and audio as input to generate gestures as a sequence of 3D poses.
As depicted in Figure \ref{fig:teaser}, it is composed of three neural networks:
\begin{enumerate}[nolistsep,noitemsep]

\item \textbf{Speech2GestExist:} A temporal CNN which takes %in 
speech as input and returns a binary flag indicating if the agent should gesture (similar to \cite{yunus2019gesture}% and to text-to-speech $f_0$ models that predict the presence of voicing separately from pitch \cite{wang2018autoregressive}%
);  
\item \textbf{Speech2GestProp:} A temporal CNN which takes %in 
speech as input and predicts a set of binary gesture properties, such as gesture type, gesture phase, etc.; 
\item \textbf{GestureFlow:} A normalizing flow \cite{kobyzev2020normalizing} that takes %in 
both speech and predicted gesture properties as input, and describe\rn{s a} probability distribution over 3D poses, from which 
\rn{motion sequences} can be sampled. %sample from.
\end{enumerate}
%Let us illustrate: When receiving the following text as input ``...and then you see a round building on the right...'' the system may first predict that there should be a gesture, and then predict that that gesture should be \textit{iconic} and should represent \textit{shape} and \textit{direction}. The gesture-generation system can then use this information to generate a gesture that is appropriate for the given context.

In this study, we experiment with the first two neural networks only. \rn{We implemented the} \emph{Speech2GestProp} and \emph{Speech2GestExist} components %was modelled as a temporal CNN based on 
using dilated CNNs. %, which takes a
\rn{Their inputs are} sequences of aligned speech text and audio frames, %as input 
and they return %a gesture binary properties
\rn{a binary vector of gesture properties} (for \textit{Speech2Prop}) or a binary flag of gesture existence (for \textit{Speech2GestExist}) %for the time-frame in the middle of the input sequence
\rn{as its} output. 
By sliding a window over the speech %text and audio 
and predicting poses, frame-by-frame properties are generated at 5 fps. Text features %used
were \rn{extracted using} DistilBERT \cite{sanh2020distilbert}. % as implemented by HuggingFace \cite{wolf-etal-2020-transformers}. 
Audio features were log\hk{-scaled} mel-spectrograms.% \rn{.} %used were spectrogram. 
%Since any given gesture property %only  is present in \hk{just} a fraction of the time frames, gesture property training data is highly imbalanced. To mitigate this\hk{,} we use \rn{a} class-balancing loss \cite{cui2019class} based on focal loss \cite{lin2017focal}.

%To mitigate class imbalance %issue 
%(any gesture property value is true only for a small fraction of the time-frames and hence most of the values for any class are zeros) we use \rn{a} class-balancing loss \cite{cui2019class} based on Focal Loss \cite{lin2017focal}. 

\section{Preliminary results}
\label{sec:results}

%To %evaluate
%\rn{explore} the feasibility of the proposed gesture-generation framework, we %built
%\rn{implemented} and evaluated %a Speech2Prop 
%\rn{the first two} component\rn{s}: \rn{the predictor of whether the agent should gesture, and} the predictor of gesture properties based on speech. 

\paragraph{Dataset}
We evaluated our model on the SaGA direction-giving dataset \cite{lucking2013data} designed to contain many representational gestures%
%and comes with detailed annotations of gesture properties. 
The dataset contains audio/video recordings of 25 participants (all German native speakers) describing the same %root to another participant.
\rn{route to other participants}
and includes detailed annotations of gesture properties.
%This dataset has a
%The limited scope of this data should make gesture modeling easier.

We considered the following three gesture properties: 1) \textit{Phase} (preparation, pre-stroke hold, stroke, post-stroke hold, and retraction); 2) \textit{Type} (deictic,	beat, iconic \cite{mcneill1992hand}, and discourse); 3) \textit{Semantic information} (amount, shape, direction, size, as described in \cite{bergmann2006verbal}).

\paragraph{Experimental Results}
For each of our experiments \rn{we calculated the mean and standard deviation of the \rn{$\text{F}_1$ score}} %we evaluated  results 
across 20-fold cross\rn{-}validation\rn{.} %and calculated \rn{$\text{F}_1$ score}
\rn{The $\text{F}_1$ score is preferable over accuracy here} since the data is highly unbalanced and accuracy does not represent \rn{overall} performance well. For gesture category and phase we report Macro F1 score  \cite{yang1999re}, since those properties are not mutually exclusive.

First we validated that gesture presence can be predicted from the speech in our dataset%:
\rn{. W}e achieved a $70\%\pm3.7\%$ Macro \rn{$\text{F}_1$} score for %predict gesture presence binary variable. This result 
binary classification, which aligns with previous work \cite{yunus2019gesture}.

\rn{Next, we experimented with predicting gesture properties.
Table \ref{tab:exp_results} contains results for predicting \rn{the} gesture category,} %(called ``Phrase'' in this dataset),
gesture semantic information, and gesture phase  from speech text and audio. 
We can see that this is a challenging task, but we are still able to predict most of the values better than chance. This was unexpected given how complex gesture semantics tend to be and could be due to the focused scope of the direction-giving task.
For a deeper study with more results and analyses, please see the follow-up work \cite{kucherenko2021multimodal}.

%\rn{Our predictions} performed well: around 60\% Macro \rn{$\text{F}_1$ score} for most classes.
%This was unexpected given how complex gesture semantics tend to be and could be due to the focused scope of the consistent-route direction-giving task.

%Our results are consistent with previous findings that gesture stroke can be predicted from audio \cite{yunus2019gesture}. The gesture preparation phase has F1 score of 0, probably because very few frames (206 out of 70k) in the dataset were labeled as being preparation phase.

\section{Discussion}

In this section we discuss the feasibility of the proposed approach. %and the limitations of the presented study.
%\subsection{Feasibility}
%\subsection{Why it should work?}
%We argue our approach is feasible because of the following reasons.
% COPIED FROM SECTION 3
% Our approach is inspired by a recent successful application of MoGlow \cite{henter2019moglow}, a Normalizing Flow based motion synthesis model, to the gesture synthesis problem. Alexanderson et al. \cite{alexanderson2020style} have shown that such models can be seamlessly conditioned on various gesture properties (height, speed, range, and similar in their case).
\rn{Our} proposal to use probabilistic models (especially normalizing flows)
\rn{is inspired by a recent application of MoGlow \cite{henter2019moglow} to perform gesture synthesis }%, was shown 
by \rn{\citeauthor{alexanderson2020style} \cite{alexanderson2020style}. They showed that such models can be seamlessly conditioned on various kinematic gesture properties (such as speed, range, and hand height), suggesting that it is possible to condition gestures on semantic properties as well.}

We obtained good results for the gesture-property prediction part of our proposed system, as described in Section \ref{sec:results}. Since we can predict several important properties with \rn{$\text{F}_1$ scores} significantly above chance level, \rn{we believe that} our predictions are reasonable and \rn{will be} useful for more appropriate gesture synthesis.

    %to be able to add style control to the gesture generation \cite{alexanderson2020style}. Similarly, as MoGlow was previously conditioned on gesture kinematic properties, it can be conditioned on gesture semantic properties.

Our two-stage approach lets the machine learning model leverage additional information (such as detailed annotation) about human gestures. It also allows direct control of gesture frequency, by adjusting the threshold on the output of \emph{Speech2GestExist} needed to trigger a gesture. Finally, it helps the model learn from small datasets, since each sub-module has a more straightforward task than learning everything at once and also can be trained separately.

%\subsection{Limitations}
%The present study has several limitations. But I am not sure if we actually have space to write about them.

\section{Conclusion}

We presented a novel gesture generation framework aiming to bridge the semantic gap between rule-based and data-driven models.\blfootnote{Authors are grateful to Stefan Kopp for providing the SaGA dataset and fruitful discussions about it and to Olga Abramov for advising on its gesture-property processing. This work was partially supported by the Swedish Foundation for Strategic Research Grant No.\ RIT15-0107 and by the Wallenberg AI, Autonomous Systems and Software Program (WASP) funded by the Knut and Alice Wallenberg Foundation.}
Our method first predicts if a gesture is appropriate for a given point in the speech and what kind of gesture is appropriate. Once this prediction is made, it is used to condition the gesture generation model. Our gesture-property prediction results are promising and indicate that the proposed approach is feasible.


%\begin{acks}
%The authors are grateful to Stefan Kopp for providing the SaGA dataset and fruitful discussions about it and to Olga Abramov for advising on dataset gesture property processing.

%This work was partially supported by the Swedish Foundation for Strategic Research Grant No.\ RIT15-0107 (EACare) and by the Wallenberg AI, Autonomous Systems and Software Program (WASP) funded by the Knut and Alice Wallenberg Foundation.
%\end{acks}

\balance

\bibliographystyle{ACM-Reference-Format}
\bibliography{ref}

\end{document}